\documentclass[12pt,preprint]{aastex}

\newcommand{\bdv}[1]{\mbox{\boldmath$#1$}}

\def\au{{\rm AU}}

\def\eff{{\rm eff}}

\def\e{{\rm E}}
\def\bpi{{\bdv\pi}}

\def\Spitzer{{\em Spitzer}}

\begin{document}
\title{{\it Spitzer} IRAC Photometry for Time Series in Crowded Fields}


\author{
S.~Calchi Novati\altaffilmark{1,2,3,a},
A.~Gould\altaffilmark{4},
J.~C.~Yee\altaffilmark{5,b},
C.~Beichman\altaffilmark{1},
G.~Bryden\altaffilmark{6},
S.~Carey\altaffilmark{7},
M.~Fausnaugh\altaffilmark{4},
B.~S.~Gaudi\altaffilmark{4},
C.~B.~Henderson\altaffilmark{6,4,c},
R.~W.~Pogge\altaffilmark{4},
Y.~Shvartzvald\altaffilmark{6,c},
B.~Wibking\altaffilmark{4},
W.~Zhu\altaffilmark{4}\\
(Spitzer team)\\
and\\
A.~Udalski\altaffilmark{8},
R.~Poleski\altaffilmark{4},
M.~Pawlak\altaffilmark{8},
M.~K.~Szyma{\'n}ski\altaffilmark{8},
J.~Skowron\altaffilmark{8},
P.~Mr{\'o}z\altaffilmark{8},
S.~Koz{\l}owski\altaffilmark{8},
{\L}.~Wyrzykowski\altaffilmark{8},
P.~Pietrukowicz\altaffilmark{8},
G.~Pietrzy{\'n}ski\altaffilmark{8},
I.~Soszy{\'n}ski\altaffilmark{8},
K.~Ulaczyk\altaffilmark{9}\\
(OGLE group)\\
}
\altaffiltext{1}{NASA Exoplanet Science Institute, MS 100-22, California Institute of Technology, Pasadena, CA 91125, USA}
\altaffiltext{2}{Dipartimento di Fisica ``E. R. Caianiello'', Universit\`a di Salerno, Via Giovanni Paolo II, 84084 Fisciano (SA), Italy}
\altaffiltext{3}{Istituto Internazionale per gli Alti Studi Scientifici (IIASS),
Via G. Pellegrino 19, 84019 Vietri Sul Mare (SA), Italy}
\altaffiltext{4}{Department of Astronomy, Ohio State University, 140 W. 18th Ave., Columbus, OH  43210, USA}
\altaffiltext{5}{Harvard-Smithsonian Center for Astrophysics, 60 Garden St., Cambridge, MA 02138, USA}
\altaffiltext{6}{Jet Propulsion Laboratory, California Institute of Technology, 4800 Oak Grove Drive, Pasadena, CA 91109, USA}
\altaffiltext{7}{{\it Spitzer}, Science Center, MS 220-6, California Institute of Technology,Pasadena, CA, USA}
\altaffiltext{8}{Warsaw University Observatory, Al.~Ujazdowskie~4, 00-478~Warszawa,Poland}
\altaffiltext{9}{Department of Physics, University of Warwick, Gibbet Hill Road, Coventry, CV4 7AL, UK}
\altaffiltext{a}{Sagan Visiting Fellow}
\altaffiltext{b}{Sagan Fellow}
\altaffiltext{c}{NASA Postdoctoral Program Fellow}


\begin{abstract}

We develop a new photometry algorithm that is optimized for {\it Spitzer}
time series in crowded fields and that is particularly adapted
to faint and/or heavily blended targets.  We apply this to the 170
targets from the 2015 {\it Spitzer} microlensing campaign and present
the results of three variants of this algorithm in an online catalog.
We present detailed accounts of the application of this algorithm to
two difficult cases, one very faint and the other very crowded.  Several
of {\it Spitzer}'s instrumental characteristics that drive the
specific features of this algorithm are shared by {\it Kepler} and
{\it WFIRST}, implying that these features may prove to be a useful
starting point for algorithms designed for microlensing campaigns
by these other missions.

\end{abstract}

\keywords{gravitational lensing: micro -- techniques: photometric}

\section{{Introduction}
\label{sec:intro}}

The IRAC camera on the {\it Spitzer} observatory has been employed
in a vast array of scientific investigations, including some that
were not imagined at the time it was launched in 2003, much less
when it was conceived in the 1980s.  Perhaps the most famous of these
is the measurement of differential exo-planet transit depth as a function of
wavelength, which requires exquisite photometric precision in order 
to probe the exo-atmospheres \citep{charbonneau05,deming05}.  These
and other applications have stimulated the development of new photometric
techniques and pipelines.

In the last two years, {\it Spitzer} has taken on a completely new
role as a ``microlens parallax satellite'' \citep{refsdal66}.
Although this application was suggested prior to launch \citep{gould99},
and even carried out for one event (OGLE-2005-SMC-001, \citealt{dong07}),
the photometric advances required to meet this role were not even
basically understood until data were collected from the 2014 ``pilot program'':
100 hours of Director's Discretionary Time that were granted 
to assess the feasibility of {\it Spitzer} microlens parallaxes.
These data revealed that while standard ({\it Spitzer} or other)
photometric packages returned reasonably good photometry
for most bright events \citep{ob140124,ob140939,21event,ob141050},
these packages usually failed, often catastrophically, for faint
events.

After 832 hours were awarded for the 2015 season (i.e., the majority
of the 38 days for which {\it Spitzer} can observe Galactic bulge
microlensing fields during microlensing season), \citet{yee15} developed
detailed protocols to optimize efficiency to achieve the program's
primary aim of measuring the ``Galactic distribution of planets''.
Of course, one of the key parameters in these protocols was an
estimate of the IRAC $3.6\,\mu$m threshold at which one could
expect reliable {\it Spitzer} photometry.  \citet{yee15} reviewed
the results from various photometric packages as applied to the 2014 
campaign to establish a threshold assuming no further improvements,
which they parameterized by a $3.6\mu$m flux 
indicator\footnote{$L_\eff = I - 0.93\,A_I - 1.3 + 0.5\Theta(I-A_I-17.2)$,
where $\Theta$ is a step function and $A_I$ is the extinction.} $L_\eff = 15.5$.
They also analyzed the reasons for the shortcomings of existing
packages when applied to this data set and reported in outline the
ongoing work (now presented in the current paper) to resolve these
shortcomings.

One of the key points made by \citet{yee15} was that the selection
of targets had to balance two critical indicators, 
encapsulated in a ``quality factor'', which sums over all observed
events, $i$,
\begin{equation}
Q = \sum_i S_i P_i
\label{eqn:quality}
\end{equation}
where $S_i$ is the planet sensitivity and $P_i$ is the probability that
the observations would return a reliable microlens parallax (which
of course requires reliable {\it Spitzer} photometry).
This implied that events with exceptional sensitivity (large $S_i$)
should be observed
despite great risk (low $P_i$) that they would be too faint by the
time {\it Spitzer} observed them, which could be from three to 10 days
after they were recognized as sensitive to planets based on their ground-based
light curves (Figure~1 of \citealt{ob140124}).

The main class of such ultra-sensitive events is high-magnification
events \citep{griest98,gould10}, which in most cases cannot be
predicted until a few days before peak.  Hence, in most cases,
by the time that {\it Spitzer} observed such a high-magnification 
event it would already
be well past peak and therefore very faint.  Acting on the basis
of Equation~(\ref{eqn:quality}), however, the {\it Spitzer} microlens
team typically attempted to observe such high magnification events
even when the expected $3.6\,\mu$m flux was well below the threshold
identified by \citet{yee15}.  In fact, seven high-magnification events 
(with peak magnifications $A>100$)
were observed, which greatly exceeded expectations.  This also greatly
increased the stakes of improving the photometric performance at faint
flux levels.

More generally, {\it Spitzer} microlens targets
can be subject to extreme crowding, which can critically impact the
photometry of these sources even when they are relatively bright.
This can in turn adversely affect a range of applications from
measuring microlens parallaxes to using the {\it Spitzer} light curves
themselves to detect planets \citep{gouldhorne}

The protocol for the target selection 
for the 2015 Spitzer campaign
is described in detail in \cite{yee15}. A key point is
that Spitzer targets can be chosen "objectively" or
"subjectively", according to which they fulfill, or not,
a set of specified criteria. This is relevant first
to establish the cadence of the observations and second to frame
the analysis of the planet sensitivity.

In Section~\ref{sec:challenges}, we review the challenges to obtaining
excellent time series photometry in crowded fields with {\it Spitzer}
using existing packages, and we outline our algorithm for meeting
these challenges.  In Section~\ref{sec:examples}, we discuss present
applications of this algorithm to two examples of the difficult
conditions encountered while pursuing the above scientific objectives.
In Section~\ref{sec:catalog}, we describe an
online catalog of photometry from our pipeline
for all of the 170 events
observed by {\it Spitzer} in the 2015 campaign.  As we note there,
it may often be possible to improve on this pipeline photometry for
individual events of special scientific interest.  The first author
should therefore be contacted regarding applications of this photometry.
Finally, in Section~\ref{sec:discuss}, we discuss some future implications
of this work.

{\section{Challenges to Crowded Field Photometry with {\it Spitzer}}
\label{sec:challenges}}

There are broadly two classes of routines applicable to time series photometry
in crowded fields using electronic detectors: 
multi-object point spread function (PSF) photometry 
(MOPSF) and difference imaging analysis (DIA).

MOPSF (e.g., DoPhot, \citealt{dophot})
was originally designed to disentangle stars in single-epoch images
of crowded fields, such as globular clusters.  Once an ensemble of 
point sources is identified, they are fit either individually, or in
groups to an ensemble of overlapping PSFs.  In principle, this routine
can be applied to a time series by simply repeating the operation
on many frames and cross-identifying point sources.  However, if the
cadence is high (so that the stars can be approximated as not moving
between epochs), then the stability of the results can be improved by
enforcing fixed positions on the stars.  This is particularly relevant
for applications such as microlensing in which one star may substantially
change brightness while the others remain roughly constant.  In that
case, if
the star positions are not held fixed, they may drift under the
influence of changing flux ratios of overlapping PSFs, including
unresolved (i.e., unmodeled) blends.

DIA (e.g., \citealt{alard98})
works by a completely different principle. No attempt is made to identify
stars.  Rather, an essentially noise-free template is created by
stacking good-seeing images.  Each image in the series is then aligned
photometrically and geometrically with the template, and the template
is convolved with a kernel to make its PSF similar to that of the current image.
Then the current image is subtracted from the convolved template.
Stars that have varied then appear as PSF bumps or divots.  Usually these
are isolated and so can easily be centroided and then multiplied by the PSF
to yield a photometric measurement.  DIA has proved so far
superior to MOPSF that the latter is rarely used for time series photometry
in crowded fields.  At the same time, however, it is obviously useless
for photometry of constant stars since these simply disappear from the
difference images.

Neither of these techniques is suitable for the analysis of faint variable
targets in crowded {\it Spitzer} images.

For the {\it Spitzer} microlensing campaign we use channel 1 (at $3.6 \mu$m)
of the IRAC camera \citep{fazio04} with mean pixel scale
$1.221^{\prime\prime}$ and mean FWHM $1.66^{\prime\prime}$, 
with the image quality being
limited primarily by the telescope optics. Given our goal of
performing point-source photometry in a crowded field, it is important
to note that the PSF is undersampled and that there is a significant
variation in sensitivity within pixels (i.e., with the ``pixel
phase'', \citealt{ingalls12}).  To address these issues the Spitzer Science Center
developed an approach based on Point Response Functions (PRFs)
specific for IRAC (a technique previously developed for the WFPC2 and
NICMOS on the {\it Hubble Space Telescope}, \citealt{lauer99,anderson00}). 
These are meant to combine information about the PSF, the detector sampling, and the
intrapixel sensitivity
variation\footnote{http://irsa.ipac.caltech.edu/data/SPITZER/docs/dataanalysistools/tools/mopex/mopexusersguide/home/}.  
The
Spitzer Science Center has also developed a specific package, MOPEX
\citep{makovoz05}, to deal with the various aspects of the
{\it Spitzer} instruments image analysis, and specifically for the
photometry of point sources for IRAC.  

We pause to
note why the standard MOPSF and DIA packages are bound to fail, at
least in the more difficult situations of low signal-to-noise
ratio (S/N) and/or crowding. DIA does not work because it implicitly assumes 
a uniform pixel response, a condition that is not met for IRAC. 
MOPSF also does not work, but not for such a fundamental reason.
The underlying logic of standard MOPSF is actually very similar to MOPEX,
but it is not built to deal 
with the PRFs.  In this sense MOPEX can be considered a specialized
variant of MOPSF.

For {\it Spitzer} microlensing observations, each ``epoch'' of a given
target is composed by six 30s exposures, each dithered by a few
arcsec.  The {\it Spitzer} pipeline (in particular, MOPEX) provides,
in addition to these single frames, also a mosaic of the dithers of a given
epoch.  We remark that a PRF analysis only makes sense when
applied to the single frame images.  This information is blurred away
in the mosaic frames, which come, on the other hand, with a higher S/N
and look similar to optical CCD images. Standard photometry
techniques, such as aperture photometry, MOPSF, and DIA can be applied to these
images, and for bright and/or isolated stars the results can be
passable to good.  However, for faint stars in more crowded situations
the results can be poor to catastrophic.

Point-source photometry requires first,
point-source extraction and second, flux determination.  In
our analysis we take advantage of two facts, which then requires us to
update and refine the MOPEX approach, accordingly.  First, we possess quite
accurate prior knowledge of essentially all significant $3.6\,\mu$m sources
on the frame.  Second, we are dealing with a time series of images for which,
as a working hypothesis, {\it one can assume that only the microlensed source 
varies}, with all other neighboring stars assumed constant,
including those that are heavily blended with the microlensed source.
(In principle, it is possible that a neighboring star can be a variable.
However, it is straightforward to identify such stars from the residuals,
and to grant them an extra flux parameter for each epoch instead of
imposing a single constant-flux parameter.)

As foreshadowed, for point-source photometry, MOPEX
works according to essentially the same principles as DoPhot, but with
the specialized feature of employing PRFs.  In essence, the
photometry procedure is a fit based on a $\chi^2$ minimization where
the signal, over the area of interest (a few pixels around each
target), is modeled as a background term plus a linear sum of the PRFs
of the different sources contributing to each pixel. There are
therefore 3 free parameters for each source: an amplitude term, which
multiplies the corresponding PRF, and a pair of coordinates specifying
the source centroid.  As a technical point we note that the PRF is
sampled at $5\times 5=25$ pixel phase positions, i.e., five steps 
across the pixel in each direction.  This is the basis for interpolating 
the PRF at each specific pixel phase\footnote{In the current analysis 
we make use of the
publicly available cold-mission PRFs. 
These are however expected to have changed since 2009
when Spitzer entered its warm-mission operational phase \citep{warm10}.
We have tested  our algorithm using a preliminary version 
of the warm-mission PRFs (J. Ingalls,
private communication) and we do not find substantial differences
in the output photometry.}.  MOPEX can perform this fitting scheme working on single
frames individually or on multiple frames simultaneously, including frames
belonging to different epochs.  From our standpoint, the drawback of
MOPEX is its rigidity: it does not allow us to
take advantage of the two aforementioned facts specific to our
problem.  In particular, the only choice permitted by MOPEX is
whether to leave the background term as a free parameter.  The flux and
positions of all the stars involved must always be fitted, which in
our case means for each individual epoch (since the microlensed source
flux changes with epoch).  For this reason PRF-fitting
MOPEX indeed works well with our data for sufficiently isolated 
and/or bright sources. However, especially for faint stars, it
sometimes cannot locate the star at all, or if it can, incorrectly
locates the centroid due to confusion of blended objects and so
applies the wrong PRF. Or, what is potentially worse, as the
microlensed source gets brighter and fainter, MOPEX shifts its
estimate of the centroid, leading to systematic errors that vary with changing
brightness.  For the 2014 campaign, MOPEX worked reasonably well for
about half of the sample.  It worked poorly for about one quarter of the
sample and failed catastrophically for the remaining quarter.  MOPEX fared
much worse on the 2015 sample because the microlensed
sources were overall substantially fainter.

From an algorithmic standpoint, the path to resolving the difficulties
faced by MOPEX is straightforward: apply the same improvements made to
DoPhot and other MOPSF packages when they graduated from single-epoch
to time-series applications.  First, one should hold
the brightness of all stars fixed from epoch to epoch 
(after establishing that they are
not in fact variable).  Second, one should (usually)
hold fixed the angular offsets between all stars from one epoch to the next.

Implementation of the first (photometric)
constraint is straightforward.  If there
are $N$ epochs, $n$ non-varying stars in the neighborhood of the lens,
and $m\geq 1$ varying stars (including the lens), then there are
a total of $n + mN + 1$ photometric parameters (including one for
the background).

However, it is not obvious how to implement the second (astrometric)
constraint, or even whether to do so.  At the outset, prior to and independently
from the photometric analysis, we build a relative astrometric solution
to link the OGLE (or in some cases, MOA) catalog 
coordinates\footnote{All {\it Spitzer} targets are selected from
ongoing microlensing events that have been publicly alerted by
either the Optical Gravitational Lens Experiment (OGLE)
or Microlensing Observations in Astrophysics
(MOA) collaborations.  Each OGLE alert is linked to a detailed 
photometric and astrometric catalog of neighboring stars.}
 to the 
{\it Spitzer} pixel coordinates, and this for each one of the $6N$ IRAC 
frames. In one case (Method~2, below) we additionally need a relative
astrometric solution between an arbitrarily chosen IRAC ``reference''
frame and the other IRAC frames.

The great majority of $3.6\,\mu$m sources in
most microlensing fields are known with good astrometric precision
from OGLE catalogs, which have much better resolution than {\it Spitzer}.
A first approach is simply to adopt these positions.
In this case, we determine the positions
of each star in each Spitzer frame, including
the microlensing source, from a local ensemble
of reference-star triangles.

Second, one can use the OGLE positions as inputs to derive the IRAC positions
from a simultaneous fit to all frames.  Then there are $(6qN + 2(n+m))$
parameters, i.e., some number $q\geq 6$ frame-transformation
parameters for each of the $6N$ images, which are applied prior
to the analysis of each image, and the two-dimensional
offsets of each star from a fiducial origin, which is held fixed for all $6N$ images. 
More specifically, to carry out the relative astrometry
we make use of the IRAF package {\texttt geomap} with a polynomial
non-linear transformation of order 5 to 8.
The typical astrometric uncertainty that we achieve is about 0.1 pixel.

Finally, one can use the OGLE positions as inputs, but derive star positions
on each frame separately.  From the standpoint of astrometry, this
is similar to what MOPEX does on a single frame
(although it still differs from MOPEX
in that it imposes a photometric constraint).  In this case, there 
are $6N \times 2(n+m)$ astrometric parameters.

In brief,
\begin{enumerate}
\item[]{Method~1: Externally determined PRF centroid.}
\item[]{Method~2: Single PRF centroid from simultaneous fit to all epochs.}
\item[]{Method~3: Individually fitted PRF centroid for each frame.}
\end{enumerate}
We find that in the majority of cases, Method~2 works best,
but Methods~1 and 3 do sometimes work better.

Specifically, if the microlensed source is extremely faint, there is almost no
information in any of the images about the location of this source,
so the OGLE information of this source position (which is very precise
because it comes from astrometry of the difference image between when 
it is magnified and at baseline) is far superior to what can be obtained
by fitting the IRAC images.  Hence, Method~1 works best for these sources.

If the microlensed source has considerable signal even in just a subset of
images, then PRF fits to its position are usually superior
to those derived from the transformation from the OGLE frame.  This is
an empirical fact.  Its cause is not completely understood but is
most likely due to a combination of errors in the frame transformations
and the different distributions of $I$ band and $3.6\mu$m light in these
crowded fields.  Hence, either Method~2 or Method~3 is better.

If the microlensed source is sometimes relatively bright and sometimes
quite faint, then Method~2 works well but Method~3
fails because the lens position is ``lost'' on the fainter images.
By the nature of the {\it Spitzer} microlensing campaign, this
is true in the majority of cases.  However, if the microlensed
source is always bright, then Method~3 can work best, probably
because it avoids the above-mentioned frame-transformation errors.

In conclusion, our approach to the photometry for the microlensing
campaign is a variant of MOPEX, a $\chi^2$ PRF-based minimization
modeling of the signal\footnote{In its present version for the fit we
  make use of MINUIT \citep{james75} within the CERNLIB package 
https://cernlib.web.cern.ch/cernlib/}, where we
have introduced the freedom to keep some parameters fixed from
epoch to epoch, while other parameters are allowed to vary.
Specifically, we always keep all other star fluxes fixed, except for those
determined to be variables.  We usually keep all of the 
vector angular separations fixed (usually at values determined from a joint
fit to the images, but sometimes by direct input from OGLE astrometry).
For a few events that remain bright throughout the {\it Spitzer} observations,
we fit for the positions in each frame separately.

In a nutshell, we take advantage of the PRF-fitting procedure and at the
same time of the specific features of microlensing and of available data.

{\section{Two Example Applications}
\label{sec:examples}}

We illustrate our algorithm with two examples, {\it Spitzer}
targets OGLE-2015-BLG-1395 and OGLE-2015-BLG-0029.  These are chosen
because they present conditions that are challenging in
substantially different ways.  

Applications of our algorithm to two other events have already been
published, the planetary event OGLE-2015-BLG-0966 \citep{ob150966},
and the binary event with a massive-remnant candidate OGLE-2015-BLG-1285
\citep{ob151285}.  We note that the case of OGLE-2015-BLG-0966 is 
particularly interesting because the early light curve yielded
satifactory photometry with Method~1, but not Method~2.  However, later,
when the event became brighter, Method~2 became much better
(although still not quite as good as Method~1) because it was
able to apply its ``knowledge'' of the microlensed source position from
the brighter images to perform precise photometry on the fainter ones.

{\subsection{OGLE-2015-BLG-1395: High-Mag From Earth,  
Ultra-Faint From {\it Spitzer}
\label{sec:ob151395}}

OGLE-2015-BLG-1395 was discovered by
the Optical Gravitational Lens Experiment (OGLE)
based on observations with the 1.4 deg$^2$ camera on its 
1.3m Warsaw Telescope at the Las Campanas Observatory in Chile
using its Early Warning System (EWS) real-time event detection
software \citep{ews1,ews2} on 19 June UT 17:56.
The {\it Spitzer} team contacted OGLE
21 hours later to get an early update on the next night of data
and based on this immediately (20 June UT 14:30)
alerted the microlensing community
that this was a high-mag event.  For results of this monitoring,
see \citet{ob151395}.  On 22 June, the event met 
the ``objective criteria'' for selection according to the protocols
of \citet{yee15} and was scheduled for observations at a cadence
of 2/day for the first week and 1/day thereafter. 
The first {\it Spitzer} observations were
three days later, HJD 7199.50, i.e., when the ground-based event was
5.3 days past peak and the event had already fallen to $I=18.6$.
In fact, when the {\it Spitzer} data are aligned to the OGLE scale
(see below) the first point is at $I=19.0$, corresponding to
$L_\eff=16.2$, roughly 0.7 mag fainter than threshold set by \citet{yee15}.
Hence, it was uncertain whether the parallax of this event could be 
recovered.

Figure~\ref{fig:prf} shows four reductions of the {\it Spitzer} data.
MOPEX clearly fails.  The reason for this failure is that it ``finds''
the source at $\sim 3$--$4^{\prime\prime}$ from its true position. Regarding
the three variants of the new reductions, Method~1 is
satisfactory,  Method~2 works best, and Method~3 fails completely.

For many years it was believed that satellite parallax measurements
required that the satellite observe the event over peak, or at least
close enough to peak to determine the satellite-based impact parameter,
$u_{0,\rm sat}$ (e.g., \citealt{gould95,gaudi97}).  This is because
satellite-based parallaxes $\bpi_\e$ are basically determined from
\begin{equation}
\bpi_\e = {\au\over D_\perp}(\Delta\tau,\Delta\beta);
\qquad \Delta\tau = {t_{0,\oplus} - t_{0,\rm sat}\over t_\e};
\qquad \Delta\beta = \pm u_{0,\oplus} - \pm u_{0,\rm sat},
\label{eqn:pieframe}
\end{equation}
where the subscripts indicate parameters as measured from Earth
and the satellite.  Here, $(t_0,u_0,t_\e)$ are the standard point-lens
microlensing parameters: time of maximum, impact parameter, and Einstein
timescale, while ${\bf D}_\perp$ is the Earth-satellite separation projected
on the sky.

However, \citet{yee15} argued that if the {\it Spitzer} source flux
$f_s$ could be independently measured (rather than fit from the {\it Spitzer}
light curve as was previously believed necessary), 
then $t_{0,\rm sat}$ and $u_{0,\rm sat}$
could be determined even if the {\it Spitzer} data covered only a 
fragment of a post-peak light curve by making use of this independent
determination of $f_s$ together with the value
of $t_\e$ measured from Earth.  From Figure~\ref{fig:prf} it is clear
that OGLE-2015-BLG-1395 provides an excellent way to test this idea.

To implement it, we first derive an estimate of the $(I_{\rm ogle}-L_{\rm spitzer})_s$
color and then incorporate this directly into a joint fit to OGLE+{\it Spitzer}
data.  The general method for estimating $(I-L)_s$ is to measure the
source color in some ground-based bands $X-I$, then to determine
an $XIL$ ($X-I$ vs $I-L$) color-color relation based on cross-identified
field stars in ground and {\it Spitzer} images, and finally to apply
this relation to the $(X-L)_s$ color to derive $(I-L)_s$.  For example,
OGLE often has sufficient $V$-band data during the event to measure
$V-I$ from the ground-based light curve.  Alternatively, if the fit
to ground-based data
shows that the source is unblended, then the baseline $(V-I)$ color
can be used as the source color.  However, in the present case, the event 
peaked too rapidly to determine $(V-I)_s$ from the normal cadence of
OGLE $V$-band data.

We therefore use the simultaneous $I$ and $H$ data taken with
the dual-channel ANDICAM camera on the 1.3m SMARTS telescope at
CTIO, which acquired substantial data on almost all 2015 {\it Spitzer}
microlensing events, primarily for exactly this purpose.
Figure~\ref{fig:ih} shows a model-independent regression
of these data, which yield $(I-H)_{s,{\rm ctio}} = -0.624\pm 0.007$.  We then
form an $IHL$ color-color diagram (Figure~\ref{fig:ihl}).  We carry out
the color-color regression on stars $-0.5<(I-H)_{\rm ctio}<+0.2$
in order to restrict the sample to warm ($T>4300\,$K) stars that 
are in the bulge (and so behind the same column of dust), and find
$I_{\rm ctio}-L_{\rm spitzer} =  1.210(I-H)_{\rm ctio} - 3.196$ and
hence $(I_{\rm ctio}-L_{\rm spitzer})_s = -3.95$.
By regression, we obtain $I_{\rm ctio}-I_{\rm ogle}=0.73$ and so finally
$(I_{\rm ogle}-L_{\rm spitzer})_s = -4.68$.  Here $L=25-2.5\log(F_{\rm spitzer})$
where $F_{\rm spitzer}$ is the instrumental flux shown in Figure~\ref{fig:prf}
(three upper panels).  We estimate an uncertainty 
of $\pm 0.1$ on this transformation.
We report below on how the final results change with the adopted error bar.
In both cases we exclude recursively the
outliers, also indicated in the figure, from the regression analysis.

We apply this color constraint to a joint fit to OGLE and {\it Spitzer}
data (Fig.~\ref{fig:lc}), wherein  
the statistical error bars are renormalized so that $\chi^2/\textrm{dof}=1$
for each observatory.
This is the standard procedure to account any possible additional
scatter due to unmodeled systematics (this issue
is further addressed in Sect.~\ref{sec:catalog}). Specifically, the size
of the error bars is about 0.08 and 0.12 mag at the bright
and the faint end, respectively.
Actually, there are four possible fits, parameterized by 
$\pm u_{0,\oplus}$ and $\pm u_{0,\rm sat}$ as indicated by 
Equation~(\ref{eqn:pieframe}).  From the present standpoint of illustrating
how {\it Spitzer} photometry is carried out and how external constraints
are applied to the light curve, $(t_0,u_0)_{\rm sat}$  are the primary 
quantities of interest for these solutions.  We find,
\begin{eqnarray}
(t_0,u_0)_{++}&=&(7197.50,+0.280);
\qquad
(t_0,u_0)_{+-}=(7197.60,+0.282)\\
(t_0,u_0)_{--}&=&(7197.48,-0.280);
\qquad
(t_0,u_0)_{-+}=(7197.60,-0.282),
\label{eqn:pairs}
\end{eqnarray}
where the first and second subscripts refer to the signs of $u_0$
for Earth and {\it Spitzer}, respectively.
That is, the naive idea that what is constrained by the {\it Spitzer}
light curve is $(t_0,|u_0|)_{\rm sat}$ is confirmed, even though the
light curve does not cover the peak.  However, this is true only
because of the color constraint.  If the uncertainty on the color constraint
is raised from 0.1 to 0.2, the above results barely change.  However,
if it is removed, then neither $t_0$ nor $u_0$ is meaningfully constrained.

These results confirm the conjecture of \citet{yee15} on the utility
of obtaining post-peak light curves from {\it Spitzer}.  In particular,
they imply that planet-sensitive high-magnification events should be
``subjectively selected'' even when it is clear that {\it Spitzer}
observations cannot begin until several days after peak.

{\subsection{OGLE-2015-BLG-0029: Bright Source in a Crowded Field}
\label{sec:ob150029}}

OGLE-2015-BLG-0029 was discovered by OGLE on 12 February 2015. On 10
May it was ``subjectively selected'' for \Spitzer\, observations with
``objective'' 1/day cadence. This was long before \Spitzer\,
observations began, which is critical for establishing the planet
sensitivity of events that do not meet objective criteria. According
to the protocols of \citet{yee15}, planets and planet sensitivity for
subjectively selected events begin with the time of selection, not
the start of \Spitzer\, observations. As it turned out, on 1 June, the
event met the objective criteria for rising events as laid out in
\citet{yee15}. Since the cadence for \Spitzer\, observations of
objectively selected events was the same as that specified on 10 May,
\Spitzer\, observations of OGLE-2015-BLG-0029 were scheduled
objectively. These observations began on 11 June, when the event first
became visible by \Spitzer, and continued until the end of the
campaign on 19 July. As specified in \citet{yee15}, a cadence of
``1/day'' determines the relative frequency with which an event should
be observed, and the actual cadence is set by this and the available
time. Hence, this event received extra observations at the beginning
and end of the campaigns, as competing events were eliminated due to
\Spitzer's Sun-exclusion angle. Altogether, it was observed at 52
epochs.

In contrast to OGLE-2015-BLG-1395, OGLE-2015-BLG-0029 is a very bright
event with $I\sim 14.5$--15. However, similar to OGLE-2015-BLG-1395, the
MOPEX pipeline basically fails to measure the light curve for this
event, albeit not catastrophically, as was the case for OGLE-2015-BLG-1395.
See Figure~\ref{fig:0029red}.
The reason that MOPEX fails is crowding, in particular
the presence of an $I=17$ neighbor at $0.97^{\prime\prime}$.  
This neighbor, which
is $\sim 5$ times fainter than the microlensed source at $3.6\mu$m
at its faintest (first) {\it Spitzer} epoch, is clearly resolved
in OGLE data but unresolved in individual IRAC images.  What makes
this case especially instructive is that the source lies almost
exactly at the first Airy null ($1.06^{\prime\prime}$), so that it is resolvable
in principle, but within one IRAC pixel ($1.2^{\prime\prime}$).  If MOPEX is
``informed'' about the presence of this source, then it assigns
the source wildly varying positions and fluxes, which impacts
the microlensing light curve catastrophically.  If MOPEX is
not "informed" (and so treats the microlensed source and the
nearby blend as having a common position), it does better
but still has more than an order of magnitude greater scatter than our new
algorithm, particularly in its Method~2 variant.

{\section{Description of Catalog}
\label{sec:catalog}}

In Table~\ref{tab:evt_param} we list the 170 events monitored in 2015.
For each, we report the event name, the coordinates, the first and last day of
observation, and the number of observed epochs. The events were chosen based
on the microlensing alerts provided by the OGLE \citep{ogleiv} and 
MOA \citep{bond04}
collaborations. The current analysis is based on the preliminary reduced
data made available by the Spitzer Science Center almost in real time (on
average, 2--3 days after the observations). The final reduction of the data
is now publicly available at the NASA/IPAC Infrared Science Database
(IRSA, http://irsa.ipac.caltech.edu/frontpage/). Besides the full table, in
the online material we provide the light curves for the full sample of
observed events, similar to Figures~\ref{fig:prf} and \ref{fig:0029red}
(excluding of the MOPEX reductions). 
Finally, an updated version of our photometry solutions is
kept at a publicly available 
website\footnote{http://www.astronomy.ohio-state.edu/Spitzer2015/}.
In particular we provide tables with the data for all
three reductions discussed in the text. The flux is expressed in arbitrary
units. In some cases, none
of the reductions look satisfactory, which may just reflect that the
absolute level of source flux variations is very low.  
The reported error is the purely statistical error from the fitting
procedure based on the IRAC uncertainty images (and so underestimates
the true uncertainties). In particular, we attribute 
any additional scatter to systematics due to 
unmodeled physics (a completely generic
feature for all photometry algorithms in crowded fields).
These must be considered case per case during
the modeling of the events.
We note that a single parameter (the ``background'')
must account for the combined effects of unresolved stars and
incompletely modeled wings of distant bright stars, in addition
to the true background.  It is therefore expected that faint stars will
sometimes take on negative flux values.  This has no impact on microlensing
analyses, which fit only the flux variations and similarly absorb
the constant-flux term into a nuisance parameter ``$f_b$''.

We emphasize that for events of special scientific interest, further work
on reductions may improve the light curve.  In particular, for events
in which the external knowledge of the source position is important,
such as OGLE-2015-BLG-1395, this can often be improved by measuring
this position from OGLE difference images.  An additional key aspect
of the analysis that may be improved on a case by case basis is the
choice of blend stars to be fitted simultaneously with the microlensed
source. An additional aspect which may become critical,
which is not taken into account in the standard pipeline 
and that requires further specific analyses, is 
saturation and nonlinearities in the response of the detector when approaching 
the pixel full well. An example
is that of the Spitzer light curve for the microlensing event
OGLE-2015-BLG-0763, 
which is presented in the online catalog without the special treatment
required to deal with saturation. Because this event is of special
interest, it was reanalyzed making use of specific procedures to 
deal with saturation and non-linearity \citep{zhu15}.

{\section{Discussion}
\label{sec:discuss}}

The principal characteristics of the data set that drove the
design of the algorithms described here will also be present in
the {\it Kepler} K2 \citep{k2} microlensing data.  These include,
\begin{enumerate}
\item[]{1) Undersampled PSF,}
\item[]{2) Non-uniform pixel response,}
\item[]{3) Many stars overlapping the microlensed source,}
\item[]{4) Most (or all) of the other stars non-variable,}
\item[]{5) External astrometric catalog at higher resolution,}
\item[]{6) Dithered exposures,}
\item[]{7) Many epochs.}
\end{enumerate}
Hence, it may be profitable to employ similar algorithms on the K2
data set.  While the first microlensing data will not be available
at least until June 2016, there are already K2 test data from other
crowded fields on which these algorithms could be developed.

All of the above characteristics, with the exception of (5) and
the partial exception of (3) apply to {\it WFIRST} 
\citep{wf1,wf2}
as well.
In some respects, the challenges of {\it WFIRST} may seem less
pressing.  However, if having available a source catalog at
higher resolution proves important for {\it WFIRST} (as it so
far appears to be for {\it Spitzer}), then now is the time to
remedy its absence, before the demise of the {\it Hubble Space Telescope}.

\acknowledgments

We thank J.\ Ingalls for a useful discussion about PRF fitting with IRAC 
data.  Work by SCN, AG, SC, JCY, and WZ was supported by JPL grant 1500811.  
Work by YS was supported by an
appointment to the NASA Postdoctoral Program at the Jet
Propulsion Laboratory, administered by Oak Ridge Associated
Universities through a contract with NASA.
Work by JCY was
performed under contract with the California Institute of Technology
(Caltech)/Jet Propulsion Laboratory (JPL) funded by NASA through the
Sagan Fellowship Program executed by the NASA Exoplanet Science
Institute.
This work is based (in part) on observations made with the 
Spitzer Space Telescope, which is operated by the Jet Propulsion Laboratory, 
California Institute of Technology under a contract with NASA. 
Support for this work was provided by NASA through an award 
issued by JPL/Caltech.

The OGLE project has received funding from the National Science Centre,
Poland, grant MAESTRO 2014/14/A/ST9/00121 to AU

\begin{figure}
\plotone{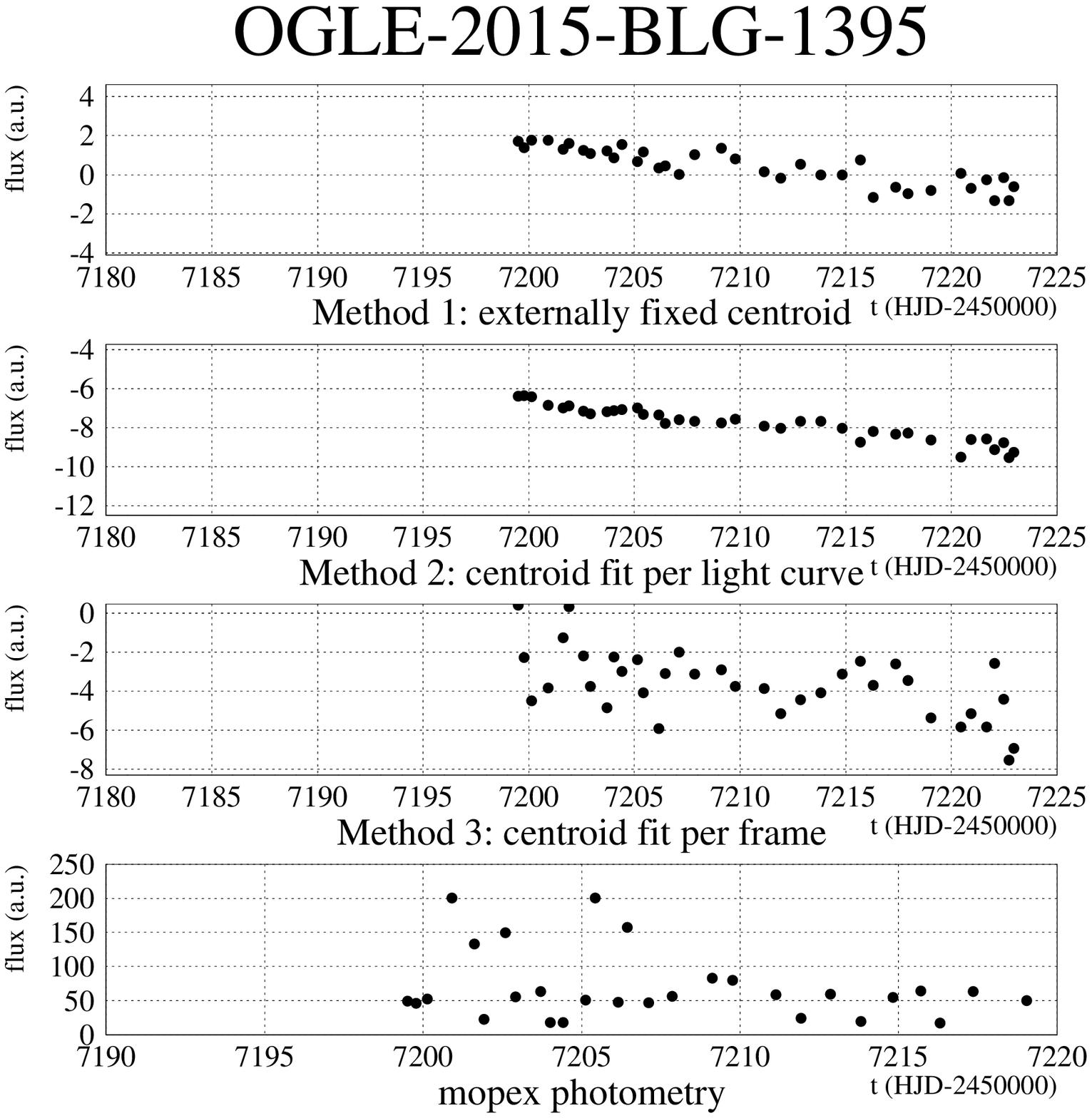}
\caption{{\it Spitzer} photometry of OGLE-2015-BLG-1395 using four different
packages: MOPEX, and the three variants of the algorithm
presented in this paper; Method~1: externally fixed centroid, 
Method~2: centroid jointly fitted to entire light curve, Method~3:
centroid fitted separately for each epoch.  Both MOPEX and Method~3
fail completely and for the same reason: very faint target.
}
\label{fig:prf}
\end{figure}

\begin{figure}
\plotone{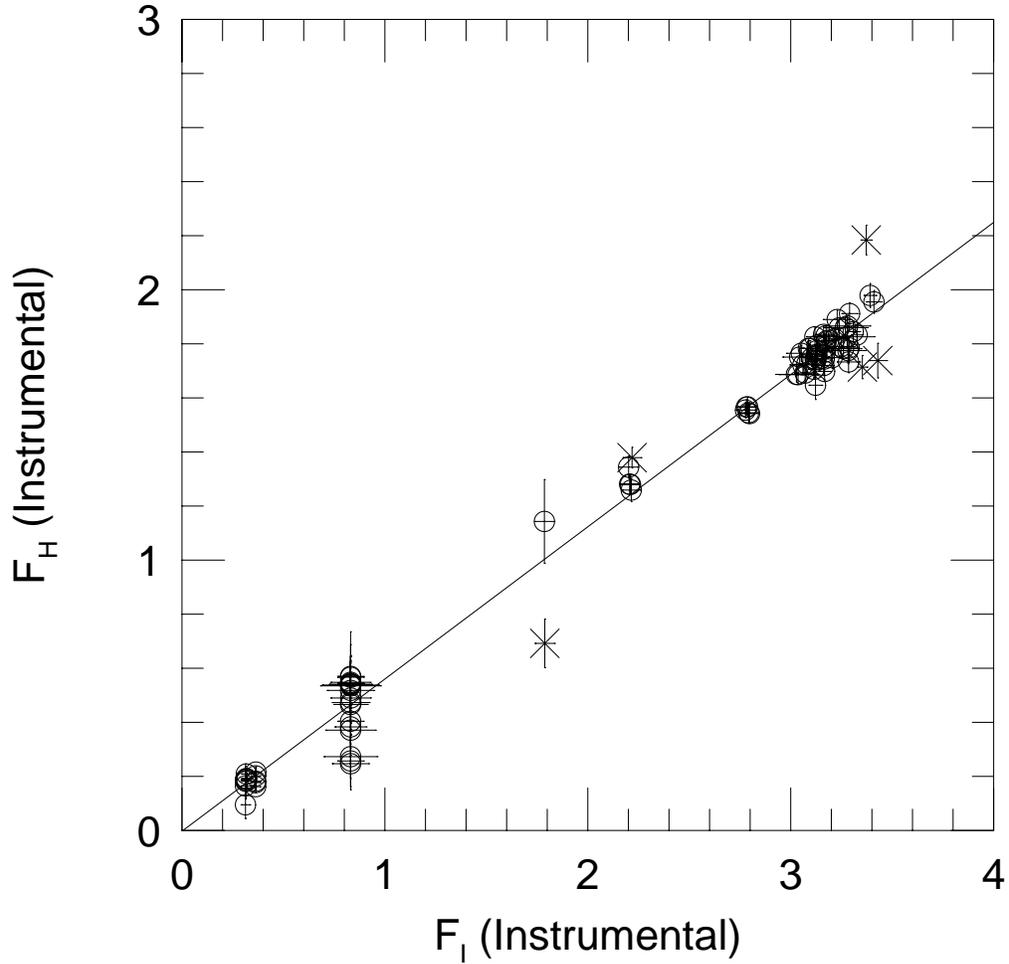}
\caption{Model-independent regression of $H$-band on $I$-band instrumental
fluxes from CTIO-SMARTS observations of OGLE-2015-BLG-1395.  These
yield a model-independent instrumental source color 
$(I-H)_{s,\rm ctio}=-2.5 \log(F_I/F_H) \approx -2.5 \log(4.00/2.25) = -0.624\pm 0.007$.
The circles indicate the points used for the color determination
whereas recursively excluded points are indicated by crosses.
}
\label{fig:ih}
\end{figure}

\begin{figure}
\plotone{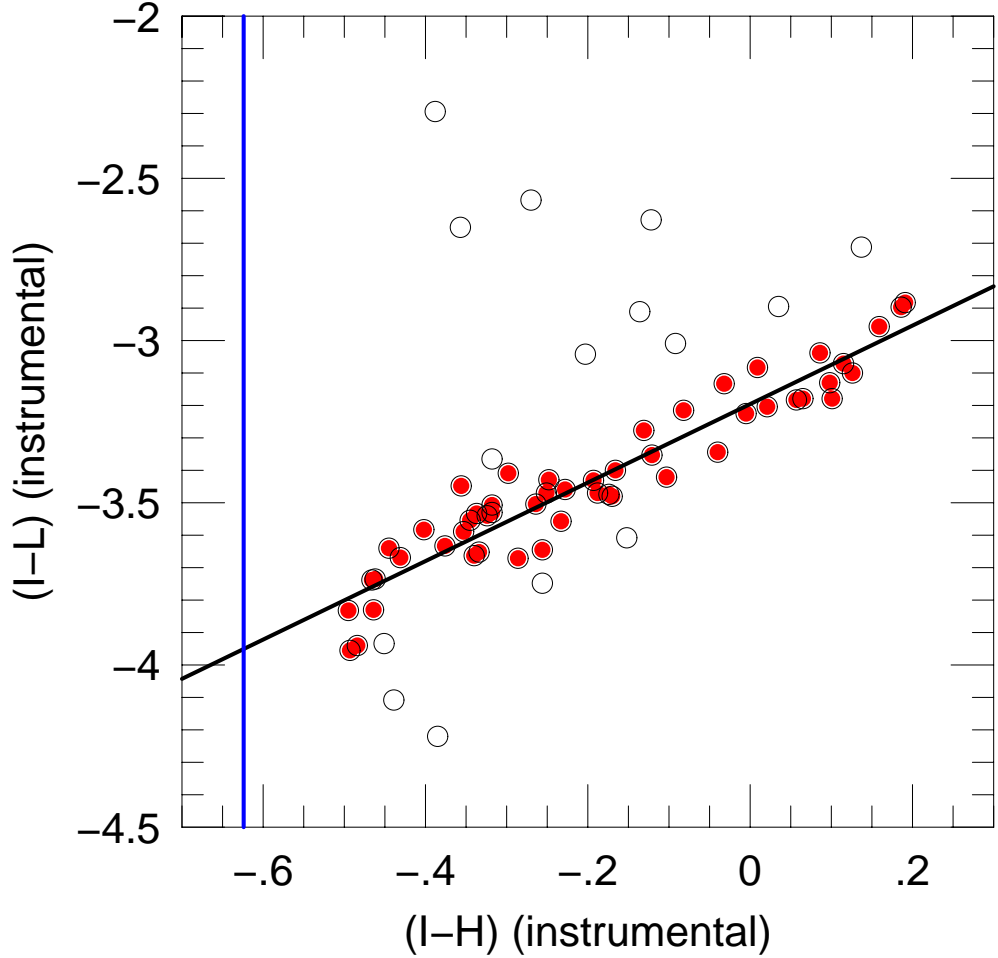}
\caption{Color-color diagram $I_{\rm ctio}-L_{\rm spitzer}$ vs.\ $(I-H)_{\rm ctio}$ 
based on field stars near OGLE-2015-BLG-1395.  Only stars
$-0.5<(I-H)_{\rm ctio}<0.2$ are included in the fit to ensure that it
is dominated by bulge stars that are reasonably warm $T>4250$.
The open points have been recursively excluded.
The blue line is the source color measured in Figure~\ref{fig:ih}.
From this we derive an instrumental color 
$(I_{\rm ctio}-L_{\rm spitzer})_s = -3.95\pm 0.10$ .
}
\label{fig:ihl}
\end{figure}

\begin{figure}
\plotone{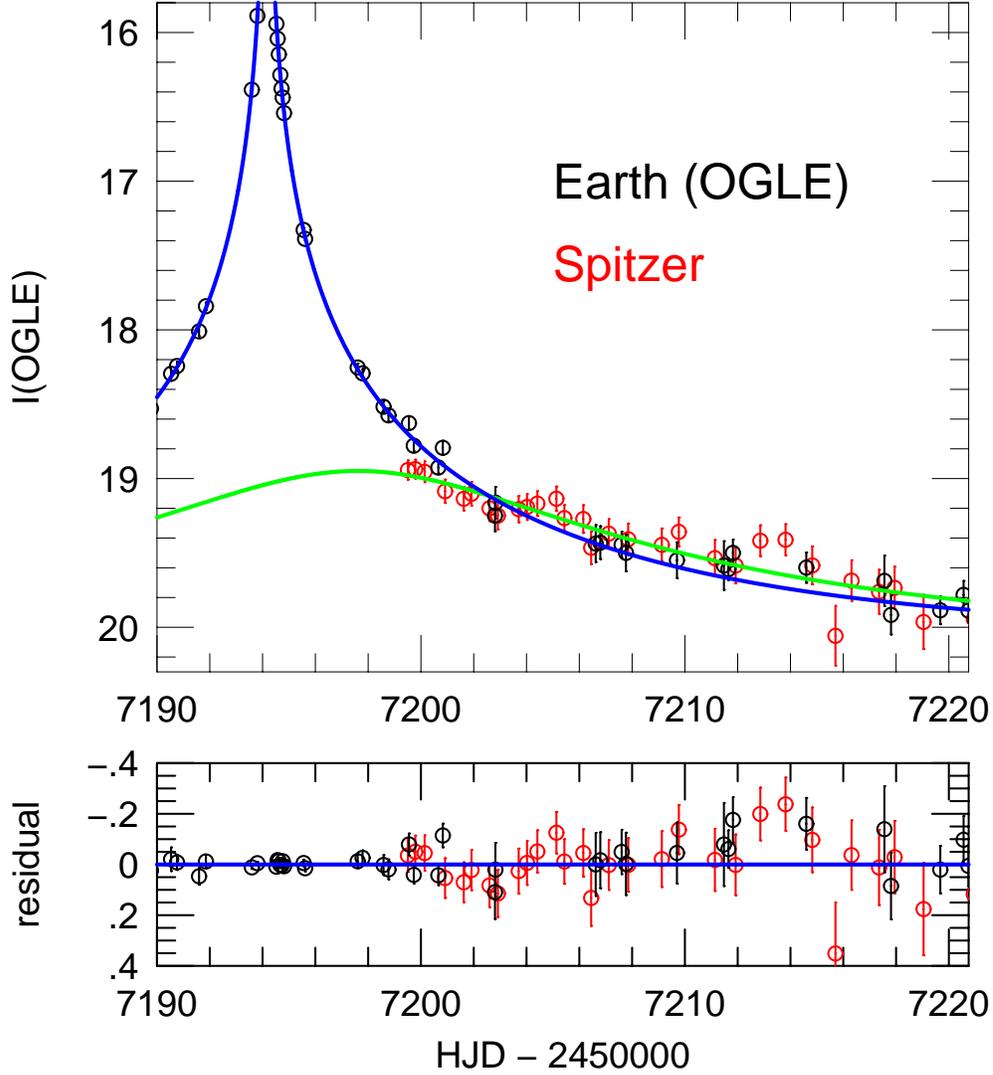}
\caption{Joint point-lens parallax fit of OGLE and {\it Spitzer}
observations of OGLE-2015-BLG-1395, using Method~2 reductions for
{\it Spitzer}.  The OGLE data (black) are shown
at their observed calibrated $I$-band magnitude.  The {\it Spitzer}
data (red) are scaled so that their ``$I$ magnitude'' corresponds
to the same magnification as an OGLE point at this same magnitude.
The fit is constrained by the
$(I_{\rm ogle}-L_{\rm spitzer})_s = -4.68\pm 0.10$ source color, which
is taken from Figure~\ref{fig:ihl} (after accounting for the
instrumental color relation $I_{\rm ctio}-I_{\rm ogle}=0.68$).
}
\label{fig:lc}
\end{figure}

\begin{figure}
\plotone{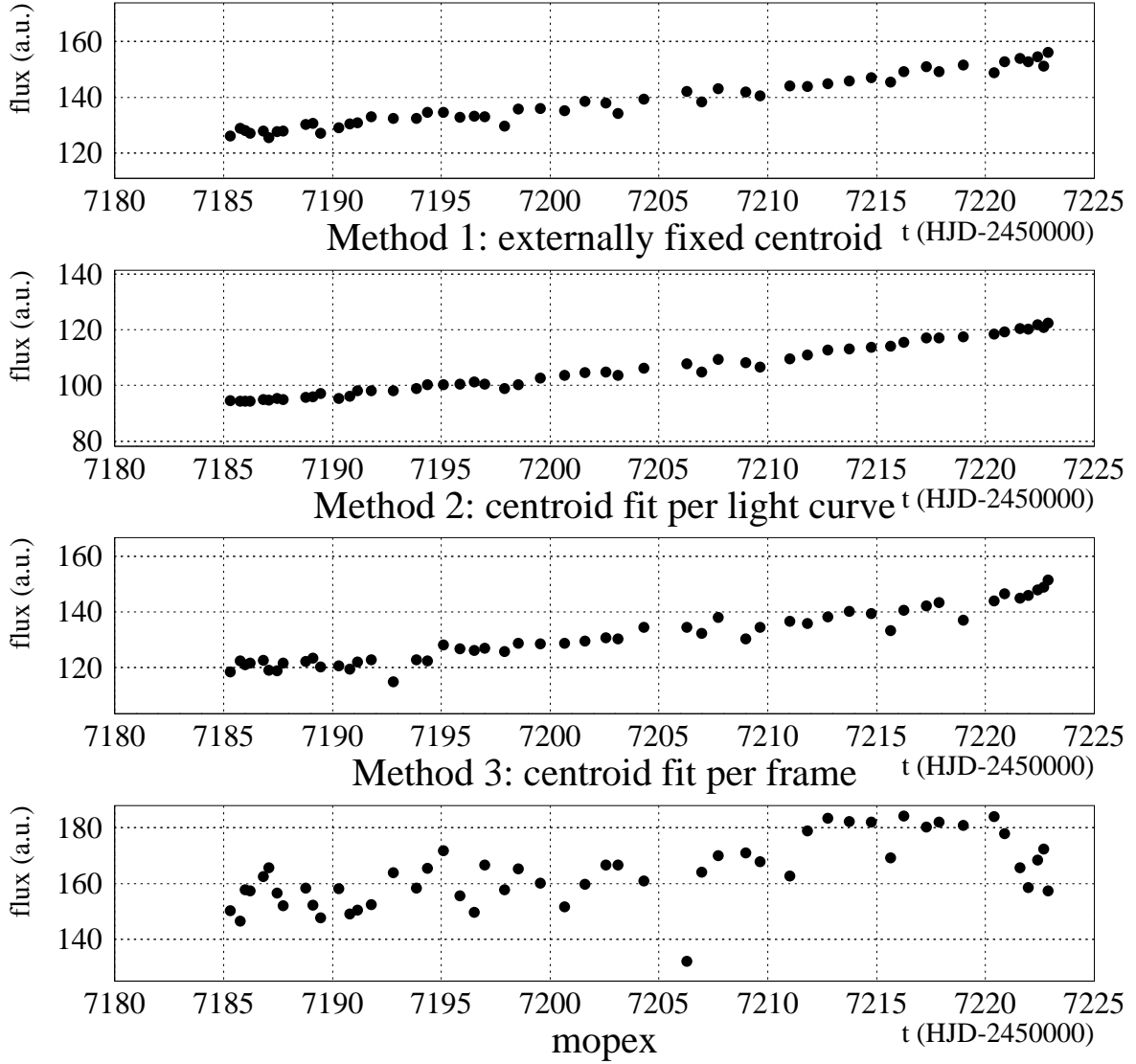}
\caption{Four reductions of OGLE-2015-BLG-0029 \Spitzer\, data. From top to bottom they are Methods~1, 2, 3, and the standard MOPEX pipeline.  Although
the microlensed source is bright, it is blended with a $\sim 5$ times
fainter star at $1^{\prime\prime}$,
which is similar to the first Airy null ($1.06^{\prime\prime}$) but smaller
than an IRAC pixel ($1.2^{\prime\prime}$).  Method~2 copes with this
situation best.  MOPEX becomes completely confused if it is ``informed''
of the existence of this blend (not shown) and does better if it
treats the blend as coincident with the microlensed source (displayed),
but the results are still quite poor when compared to Method~2.}
\label{fig:0029red}
\end{figure}

\begin{deluxetable}{lccccr}
\tabletypesize{\scriptsize}
\tablecaption{Observed Events\label{tab:evt_param}}
\tablewidth{0pt}
\tablehead{
\colhead{Event} & \colhead{RA (J2000)} &\colhead{DEC (J2000)}&
\multicolumn{3}{c}{Spitzer}\\
\cline{4-6}\\
\colhead{} & \colhead{} &\colhead{}&
\colhead{first}&\colhead{last}&\colhead{epochs}\\
& (degrees) & (degrees) & (HJD)& (HJD)&
}
\startdata
 MOA-BLG-2015-020   & 268.219917 & -32.485861   & 7183.95 & 7221.76 & 61\\     
 MOA-BLG-2015-079   & 268.085334 & -31.295665   & 7183.66 & 7221.76 &112\\
 MOA-BLG-2015-117   & 268.792652 & -35.080247   & 7184.10 & 7212.60 & 45\\ 
 MOA-BLG-2015-220   & 271.832485 & -25.467555   & 7187.12 & 7222.96 & 48\\   
 MOA-BLG-2015-237   & 269.467804 & -31.472354   & 7184.97 & 7223.06 & 54\\   
 MOA-BLG-2015-240   & 270.585946 & -31.996441   & 7185.80 & 7212.75 & 37\\   
 MOA-BLG-2015-245   & 269.321669 & -27.857411   & 7184.99 & 7222.60 & 78\\   
 MOA-BLG-2015-267   & 269.214339 & -30.775359   & 7199.49 & 7222.65 & 39\\   
 MOA-BLG-2015-282   & 272.572708 & -29.265556   & 7199.71 & 7222.94 & 29\\   
 MOA-BLG-2015-287   & 270.125760 & -33.017623   & 7199.35 & 7222.89 & 36\\   
 OGLE-BLG-2014-0298 & 268.565625 & -31.018083   & 7185.73 & 7221.84 & 25\\     
 OGLE-BLG-2014-0613 & 268.490300 & -28.572670   & 7183.98 & 7221.81 &158\\     
 OGLE-BLG-2015-0011 & 269.217833 & -29.283250   & 7184.96 & 7222.59 & 53\\     
 OGLE-BLG-2015-0022 & 268.090792 & -29.785028   & 7183.67 & 7221.05 & 60\\     
 OGLE-BLG-2015-0029 & 269.944167 & -28.644944   & 7185.31 & 7222.89 & 52\\     
 OGLE-BLG-2015-0034 & 270.580333 & -27.516083   & 7186.01 & 7222.92 & 62\\     
 OGLE-BLG-2015-0060 & 269.993125 & -27.780944   & 7185.78 & 7222.90 & 50\\     
 OGLE-BLG-2015-0081 & 268.653000 & -28.996278   & 7184.10 & 7221.82 & 57\\     
 OGLE-BLG-2015-0145 & 270.177833 & -35.154056   & 7185.29 & 7222.88 & 51\\     
 OGLE-BLG-2015-0149 & 270.288208 & -32.557694   & 7185.80 & 7222.93 & 49\\     
 OGLE-BLG-2015-0195 & 266.790792 & -33.417278   & 7182.39 & 7220.58 & 70\\     
 OGLE-BLG-2015-0196 & 266.492958 & -32.956778   & 7182.39 & 7220.56 & 70\\     
 OGLE-BLG-2015-0211 & 262.359083 & -30.981750   & 7180.20 & 7216.80 &109\\     
 OGLE-BLG-2015-0238 & 265.708250 & -36.818500   & 7182.37 & 7219.17 & 75\\     
 OGLE-BLG-2015-0244 & 268.240625 & -28.980889   & 7183.97 & 7221.75 & 59\\     
 OGLE-BLG-2015-0254 & 271.553292 & -26.850194   & 7186.87 & 7222.96 & 49\\     
 OGLE-BLG-2015-0326 & 274.074625 & -26.338139   & 7199.73 & 7223.03 & 29\\     
 OGLE-BLG-2015-0350 & 268.248583 & -31.820278   & 7183.95 & 7221.76 & 61\\     
 OGLE-BLG-2015-0379 & 269.104292 & -29.574056   & 7184.61 & 7222.58 & 54\\     
 OGLE-BLG-2015-0388 & 268.468917 & -28.534028   & 7183.98 & 7221.81 & 66\\     
 OGLE-BLG-2015-0401 & 264.081167 & -27.774139   & 7180.19 & 7213.03 & 94\\     
 OGLE-BLG-2015-0444 & 262.917375 & -30.673111   & 7180.20 & 7216.79 & 84\\     
 OGLE-BLG-2015-0448 & 272.559917 & -31.752611   & 7187.54 & 7223.04 &211\\     
 OGLE-BLG-2015-0461 & 270.043208 & -28.156944   & 7185.79 & 7222.90 & 58\\     
 OGLE-BLG-2015-0477 & 273.200500 & -25.014028   & 7188.88 & 7222.97 & 45\\     
 OGLE-BLG-2015-0479 & 265.918958 & -35.509278   & 7182.38 & 7219.17 & 67\\     
 OGLE-BLG-2015-0529 & 270.264667 & -29.922917   & 7185.79 & 7222.92 & 51\\     
 OGLE-BLG-2015-0561 & 267.224542 & -29.674444   & 7183.09 & 7220.61 & 68\\     
 OGLE-BLG-2015-0565 & 269.153708 & -29.128056   & 7184.62 & 7222.59 & 53\\     
 OGLE-BLG-2015-0572 & 272.847750 & -25.473861   & 7188.88 & 7222.97 & 44\\     
 OGLE-BLG-2015-0607 & 274.049625 & -27.451306   & 7189.13 & 7223.03 & 54\\     
 OGLE-BLG-2015-0642 & 263.539458 & -24.882528   & 7192.99 & 7216.78 & 25\\     
 OGLE-BLG-2015-0692 & 268.079708 & -28.133917   & 7185.70 & 7221.07 & 73\\     
 OGLE-BLG-2015-0703 & 269.048583 & -29.853389   & 7184.47 & 7222.17 & 65\\     
 OGLE-BLG-2015-0709 & 266.113292 & -34.442972   & 7182.38 & 7219.16 & 74\\     
 OGLE-BLG-2015-0713 & 272.660042 & -28.412083   & 7213.65 & 7223.01 & 28\\     
 OGLE-BLG-2015-0749 & 264.955708 & -27.511722   & 7180.78 & 7206.48 & 59\\     
 OGLE-BLG-2015-0761 & 272.084667 & -25.443056   & 7187.52 & 7222.97 & 46\\     
 OGLE-BLG-2015-0763 & 263.097542 & -29.302694   & 7180.19 & 7216.78 &142\\
 OGLE-BLG-2015-0769 & 265.670083 & -27.368833   & 7182.40 & 7219.15 & 59\\     
 OGLE-BLG-2015-0772 & 269.421583 & -28.143139   & 7184.99 & 7222.60 & 54\\     
 OGLE-BLG-2015-0798 & 271.680583 & -27.590167   & 7186.86 & 7222.95 & 47\\     
 OGLE-BLG-2015-0802 & 266.514625 & -36.497917   & 7182.38 & 7220.57 & 76\\     
 OGLE-BLG-2015-0808 & 267.184833 & -33.102278   & 7199.31 & 7220.59 & 25\\     
 OGLE-BLG-2015-0828 & 267.132917 & -34.184778   & 7183.08 & 7220.58 & 72\\     
 OGLE-BLG-2015-0843 & 271.197042 & -27.176472   & 7186.87 & 7222.95 & 61\\     
 OGLE-BLG-2015-0845 & 271.088708 & -31.580556   & 7186.24 & 7222.93 &148\\     
 OGLE-BLG-2015-0877 & 273.676958 & -27.970917   & 7188.92 & 7212.92 & 29\\     
 OGLE-BLG-2015-0912 & 265.957583 & -26.519944   & 7182.41 & 7219.15 & 57\\     
 OGLE-BLG-2015-0914 & 269.311750 & -29.760139   & 7184.96 & 7222.58 & 53\\     
 OGLE-BLG-2015-0925 & 270.259292 & -26.384889   & 7199.67 & 7222.91 & 29\\     
 OGLE-BLG-2015-0926 & 270.883792 & -26.135806   & 7199.67 & 7222.91 & 29\\     
 OGLE-BLG-2015-0930 & 264.393917 & -25.512667   & 7180.21 & 7218.07 &111\\     
 OGLE-BLG-2015-0941 & 272.348083 & -25.762056   & 7187.53 & 7222.98 & 47\\     
 OGLE-BLG-2015-0944 & 265.258000 & -27.853806   & 7181.09 & 7198.12 & 42\\     
 OGLE-BLG-2015-0955 & 273.709833 & -27.006833   & 7213.88 & 7223.01 & 14\\     
 OGLE-BLG-2015-0958 & 267.754958 & -28.416861   & 7183.32 & 7221.06 & 64\\     
 OGLE-BLG-2015-0961 & 268.146375 & -30.080361   & 7185.69 & 7221.75 &131\\     
 OGLE-BLG-2015-0965 & 269.291083 & -28.961056   & 7184.99 & 7222.59 & 52\\     
 OGLE-BLG-2015-0966 & 268.754250 & -29.047111   & 7192.64 & 7222.14 &119\\     
 OGLE-BLG-2015-0968 & 268.985833 & -28.854083   & 7184.47 & 7222.19 & 56\\     
 OGLE-BLG-2015-0977 & 272.061375 & -26.159472   & 7187.53 & 7222.96 & 47\\     
 OGLE-BLG-2015-0984 & 263.875167 & -26.553778   & 7185.84 & 7217.51 & 39\\     
 OGLE-BLG-2015-0987 & 268.193708 & -28.361222   & 7183.97 & 7221.07 & 65\\     
 OGLE-BLG-2015-1020 & 263.050417 & -29.781722   & 7180.19 & 7216.78 &108\\     
 OGLE-BLG-2015-1076 & 270.744792 & -32.853694   & 7192.81 & 7222.94 &135\\     
 OGLE-BLG-2015-1077 & 274.478083 & -22.992611   & 7192.93 & 7223.03 & 64\\     
 OGLE-BLG-2015-1084 & 271.146417 & -26.168028   & 7186.87 & 7212.80 & 35\\     
 OGLE-BLG-2015-1095 & 267.693875 & -32.708167   & 7183.24 & 7221.04 & 65\\     
 OGLE-BLG-2015-1096 & 267.872333 & -32.153694   & 7183.40 & 7221.04 & 62\\     
 OGLE-BLG-2015-1100 & 263.669042 & -29.885194   & 7180.20 & 7217.52 & 84\\     
 OGLE-BLG-2015-1109 & 270.213083 & -30.645139   & 7185.80 & 7191.80 & 13\\     
 OGLE-BLG-2015-1112 & 273.535250 & -25.482778   & 7188.79 & 7222.98 & 51\\     
 OGLE-BLG-2015-1113 & 264.277208 & -27.299083   & 7180.18 & 7218.10 & 88\\     
 OGLE-BLG-2015-1123 & 273.648042 & -26.507389   & 7188.88 & 7223.01 & 70\\     
 OGLE-BLG-2015-1124 & 273.253750 & -27.532861   & 7188.89 & 7206.39 & 23\\     
 OGLE-BLG-2015-1129 & 266.413125 & -27.160028   & 7182.41 & 7185.33 & 18\\     
 OGLE-BLG-2015-1136 & 268.557583 & -28.708361   & 7185.71 & 7198.40 & 24\\     
 OGLE-BLG-2015-1145 & 269.451042 & -29.080444   & 7185.63 & 7222.84 & 79\\     
 OGLE-BLG-2015-1148 & 270.259625 & -28.686111   & 7185.79 & 7198.57 & 22\\     
 OGLE-BLG-2015-1150 & 273.049208 & -25.661694   & 7199.72 & 7223.05 &121\\     
 OGLE-BLG-2015-1153 & 264.317792 & -29.426389   & 7193.01 & 7218.08 & 50\\     
 OGLE-BLG-2015-1161 & 267.618375 & -29.979222   & 7192.59 & 7220.61 & 34\\     
 OGLE-BLG-2015-1167 & 269.024500 & -28.626361   & 7184.46 & 7222.20 & 56\\     
 OGLE-BLG-2015-1172 & 268.545417 & -31.103194   & 7185.73 & 7198.41 & 24\\     
 OGLE-BLG-2015-1184 & 268.514458 & -30.311250   & 7183.96 & 7185.41 & 10\\     
 OGLE-BLG-2015-1187 & 268.657792 & -31.681222   & 7199.45 & 7222.16 & 27\\     
 OGLE-BLG-2015-1188 & 268.877125 & -31.281667   & 7184.34 & 7222.17 & 56\\     
 OGLE-BLG-2015-1189 & 270.735000 & -30.373056   & 7192.80 & 7222.93 & 87\\     
 OGLE-BLG-2015-1194 & 261.049625 & -28.947472   & 7207.44 & 7215.02 &  9\\     
 OGLE-BLG-2015-1196 & 266.466667 & -24.287278   & 7182.41 & 7198.06 & 37\\     
 OGLE-BLG-2015-1197 & 264.229917 & -29.782028   & 7207.42 & 7218.09 & 31\\     
 OGLE-BLG-2015-1204 & 268.927083 & -29.384306   & 7199.50 & 7222.18 & 27\\     
 OGLE-BLG-2015-1207 & 267.709667 & -23.445500   & 7192.93 & 7220.56 & 31\\     
 OGLE-BLG-2015-1212 & 268.103292 & -29.181111   & 7185.70 & 7206.09 & 92\\     
 OGLE-BLG-2015-1221 & 268.418792 & -32.279306   & 7199.33 & 7221.94 &101\\     
 OGLE-BLG-2015-1223 & 265.690333 & -27.528333   & 7182.40 & 7212.95 & 53\\     
 OGLE-BLG-2015-1227 & 268.327417 & -29.652833   & 7199.41 & 7221.83 & 26\\     
 OGLE-BLG-2015-1232 & 267.641125 & -32.739639   & 7185.66 & 7191.98 & 16\\     
 OGLE-BLG-2015-1234 & 267.179917 & -24.520028   & 7192.93 & 7220.56 & 31\\     
 OGLE-BLG-2015-1236 & 265.421375 & -27.809278   & 7185.81 & 7219.15 & 42\\     
 OGLE-BLG-2015-1241\tablenotemark{a} & 269.529708 & -27.932000   & 7185.63 & 7198.49 & 35\\     
 OGLE-BLG-2015-1256 & 269.537083 & -28.769083   & 7185.77 & 7206.27 & 31\\     
 OGLE-BLG-2015-1262 & 267.523792 & -30.712611   & 7185.68 & 7206.04 & 33\\     
 OGLE-BLG-2015-1263 & 267.717208 & -32.380389   & 7185.67 & 7206.03 & 32\\     
 OGLE-BLG-2015-1264 & 264.976250 & -34.584722   & 7192.51 & 7219.17 & 31\\     
 OGLE-BLG-2015-1268 & 269.207375 & -21.899333   & 7192.61 & 7212.93 & 48\\     
 OGLE-BLG-2015-1281 & 273.034625 & -27.167694   & 7192.86 & 7222.95 & 36\\     
 OGLE-BLG-2015-1285 & 264.846875 & -27.820278   & 7199.81 & 7218.09 & 39\\     
 OGLE-BLG-2015-1288 & 267.516625 & -33.368472   & 7192.52 & 7221.03 & 35\\     
 OGLE-BLG-2015-1289 & 269.461583 & -28.217306   & 7206.94 & 7222.83 & 22\\     
 OGLE-BLG-2015-1295 & 270.226042 & -27.302139   & 7199.39 & 7223.08 &127\\     
 OGLE-BLG-2015-1297 & 268.087542 & -29.683806   & 7192.59 & 7206.05 & 17\\     
 OGLE-BLG-2015-1303 & 273.970333 & -30.373194   & 7199.42 & 7223.02 & 68\\     
 OGLE-BLG-2015-1309 & 264.693250 & -27.479472   & 7199.82 & 7218.10 & 20\\     
 OGLE-BLG-2015-1315 & 266.672583 & -33.458194   & 7206.71 & 7220.57 & 17\\     
 OGLE-BLG-2015-1319 & 269.443333 & -32.472194   & 7206.95 & 7223.07 & 22\\     
 OGLE-BLG-2015-1325 & 266.423625 & -35.957278   & 7213.41 & 7220.83 & 18\\     
 OGLE-BLG-2015-1328 & 267.063208 & -32.687778   & 7199.32 & 7206.02 &  8\\     
 OGLE-BLG-2015-1339 & 263.321917 & -30.271222   & 7199.91 & 7217.52 & 18\\     
 OGLE-BLG-2015-1341 & 268.563667 & -28.350167   & 7199.47 & 7221.81 & 25\\     
 OGLE-BLG-2015-1344 & 268.934542 & -27.940028   & 7192.67 & 7222.32 & 57\\     
 OGLE-BLG-2015-1346 & 269.806958 & -28.936611   & 7192.79 & 7222.88 & 37\\     
 OGLE-BLG-2015-1348 & 269.487708 & -30.509417   & 7199.55 & 7223.06 & 29\\     
 OGLE-BLG-2015-1350 & 266.685583 & -36.592889   & 7199.29 & 7220.57 & 73\\     
 OGLE-BLG-2015-1352 & 265.141875 & -27.338306   & 7207.37 & 7218.08 & 13\\     
 OGLE-BLG-2015-1366 & 268.289250 & -21.944056   & 7213.90 & 7221.74 & 10\\     
 OGLE-BLG-2015-1368 & 267.945083 & -29.575167   & 7199.34 & 7221.06 & 77\\     
 OGLE-BLG-2015-1370 & 268.956667 & -29.124111   & 7206.90 & 7222.18 & 19\\     
 OGLE-BLG-2015-1371 & 268.665875 & -29.093944   & 7206.87 & 7221.83 & 18\\     
 OGLE-BLG-2015-1374 & 271.134375 & -26.887806   & 7213.64 & 7223.00 & 28\\     
 OGLE-BLG-2015-1378 & 264.834792 & -22.976222   & 7199.79 & 7218.06 & 20\\     
 OGLE-BLG-2015-1383 & 268.651292 & -28.702250   & 7213.53 & 7221.94 & 22\\     
 OGLE-BLG-2015-1392 & 264.095125 & -27.298917   & 7213.68 & 7217.82 & 11\\     
 OGLE-BLG-2015-1395 & 271.783917 & -26.303333   & 7199.51 & 7222.96 & 36\\     
 OGLE-BLG-2015-1400 & 268.109625 & -28.866250   & 7213.47 & 7221.55 & 20\\     
 OGLE-BLG-2015-1403 & 267.882875 & -29.340361   & 7206.79 & 7221.06 & 17\\     
 OGLE-BLG-2015-1412 & 270.124875 & -27.435361   & 7207.00 & 7223.05 & 35\\     
 OGLE-BLG-2015-1416 & 270.937500 & -26.848056   & 7207.07 & 7222.92 & 21\\     
 OGLE-BLG-2015-1420 & 271.477500 & -28.557361   & 7213.65 & 7223.02 & 28\\     
 OGLE-BLG-2015-1424 & 264.063292 & -29.642833   & 7199.83 & 7218.09 & 20\\     
 OGLE-BLG-2015-1425 & 263.142958 & -30.716333   & 7213.69 & 7217.10 &  9\\     
 OGLE-BLG-2015-1435 & 269.842167 & -27.219222   & 7199.57 & 7206.27 &  8\\     
 OGLE-BLG-2015-1440 & 270.973167 & -27.707417   & 7213.64 & 7222.99 & 28\\     
 OGLE-BLG-2015-1447 & 269.233417 & -31.941111   & 7213.50 & 7222.81 & 27\\     
 OGLE-BLG-2015-1448 & 270.939333 & -27.852194   & 7207.04 & 7222.99 & 35\\     
 OGLE-BLG-2015-1450 & 270.210000 & -28.671028   & 7213.51 & 7223.07 & 56\\     
 OGLE-BLG-2015-1457 & 267.857167 & -30.436389   & 7206.78 & 7221.05 & 17\\     
 OGLE-BLG-2015-1467 & 265.107500 & -23.801389   & 7207.38 & 7212.97 &  7\\     
 OGLE-BLG-2015-1469 & 265.266583 & -24.848944   & 7213.67 & 7218.92 & 13\\     
 OGLE-BLG-2015-1470 & 268.501833 & -31.592806   & 7206.82 & 7221.84 & 18\\     
 OGLE-BLG-2015-1481 & 268.008083 & -30.985028   & 7213.48 & 7221.54 & 20\\     
 OGLE-BLG-2015-1482 & 267.630542 & -30.888694   & 7206.73 & 7221.04 & 17\\     
 OGLE-BLG-2015-1485 & 265.380250 & -24.214972   & 7207.35 & 7218.92 & 20\\     
 OGLE-BLG-2015-1492 & 268.686000 & -29.707361   & 7213.48 & 7221.96 & 22\\     
 OGLE-BLG-2015-1500 & 272.021583 & -27.445611   & 7207.12 & 7212.84 &  7\\     
 OGLE-BLG-2015-1521 & 267.485792 & -21.411889   & 7213.66 & 7220.82 & 17\\     
 OGLE-BLG-2015-1530 & 269.018958 & -28.240472   & 7213.53 & 7222.38 & 24\\     
 OGLE-BLG-2015-1533 & 269.048083 & -29.476417   & 7213.51 & 7222.39 & 24\\     
 OGLE-BLG-2015-1553 & 265.857042 & -23.377750   & 7213.67 & 7219.14 & 14\\     
\enddata
\tablecomments{We report name, target coordinates and information
about Spitzer observations for the 170 microlensing events
monitored during the 2015
Spitzer microlensing campaign. We note that
2 out of the 10 MOA events have also been observed
by the OGLE collaboration (specifically
MOA-BLG-2015-020 and MOA-BLG-2015-245
are also known as OGLE-BLG-2015-0102 and 
OGLE-BLG-2015-1286, respectively) and
64 out of the 160 OGLE events
have also been observed by the MOA collaboration
(for the detail of this we refer to the MOA microlensing
alert webpage 
http://www.phys.canterbury.ac.nz/moa/microlensing\_alerts.html).
In Fig.~6 we show the corresponding lightcurves for the 3 variants
of our photometry reduction for all the events (online material).
The first method, which relies on an analysis
based on the OGLE online catalog of nearby 
stars to the microlensed source, is not presented
for MOA-only events and for 4 OGLE events
(OGLE-BLG-2015-0561, OGLE-BLG-2015-0941,
OGLE-BLG-2015-1112 and OGLE-BLG-2015-1234).
}
\tablenotetext{a}{This event is also known as OGLE-BLG-2015-1253.}
\end{deluxetable}

\end{document}